\documentclass[preprintnumbers, floatfix, preprintnumbers, letterpaper, superscriptaddress,nofootinbib, twocolumn]{revtex4}
\pdfoutput=1
\usepackage{graphicx}
\usepackage{microtype}
\usepackage{amsmath}
\usepackage{amssymb}
\usepackage{subfigure}
\usepackage{hyperref}
\usepackage{url}
\usepackage{xcolor}
\usepackage{color}
\usepackage{mathrsfs}
\usepackage{calrsfs}
\usepackage{amsfonts}
\usepackage{latexsym}
\usepackage{ragged2e}
\usepackage{epsfig}
\usepackage{textcomp}
\usepackage{phaistos}
\makeatletter
\renewcommand\@makefnmark{\hbox{\@textsuperscript{\normalfont\color{purple}\@thefnmark}}}
\renewcommand\@makefntext[1]{%
  \parindent 1em\noindent
            \hb@xt@1.8em{%
                \hss\@textsuperscript{\normalfont\@thefnmark}}#1}
\makeatother

\usepackage{caption}
\DeclareCaptionJustification{justified}{\leftskip=0pt \rightskip=0pt \parfillskip=0pt plus 1fil}
\definecolor{vividviolet}{rgb}{0.62, 0.0, 1.0}
\definecolor{amaranth}{rgb}{0.9, 0.17, 0.31}
\definecolor{palatinateblue}{rgb}{0.15, 0.23, 0.89}
\definecolor{brightpink}{rgb}{1.0, 0.0, 0.5}
\definecolor{cornflowerblue}{rgb}{0.39, 0.58, 0.93}
\definecolor{deepcarminepink}{rgb}{0.94, 0.19, 0.22}
\definecolor{radicalred}{rgb}{1.0, 0.21, 0.37}

\hypersetup{ linktoc=all,
    colorlinks, linkcolor={palatinateblue},
    citecolor={brightpink}, urlcolor={amaranth}
}

\graphicspath{{Images/}}

\renewcommand{\d}[1]{\ensuremath{\operatorname{d}\!{#1}}}

\graphicspath{{Images/}}
\renewcommand{\d}[1]{\ensuremath{\operatorname{d}\!{#1}}}
\makeatletter
\def\@fnsymbol#1{\ensuremath{\ifcase#1\or $\textleaf$ \or $\PHplaneTree$
\else\@ctrerr\fi}}%

\makeatother

\def\sideremark#1{\ifvmode\leavevmode\fi\vadjust{\vbox to0pt{\vss
 \hbox to 0pt{\hskip\hsize\hskip1em
 \vbox{\hsize1.5cm\tiny\raggedright\pretolerance10000
 \noindent #1\hfill}\hss}\vbox to8pt{\vfil}\vss}}}%
\begin{document}

\title{Charged Particle Production Rate from Cosmic Censorship \\in Dilaton Black Hole Spacetimes}

\author{Yen Chin \surname{Ong}}
\email{ycong@yzu.edu.cn}
\affiliation{Center for Gravitation and Cosmology, College of Physical Science and Technology, Yangzhou University, \\180 Siwangting Road, Yangzhou City, Jiangsu Province 225002, China}
\affiliation{School of Aeronautics and Astronautics, Shanghai Jiao Tong University, \\800 Dongchuan Road, Minhang District, Shanghai 200240, China
}

\author{Yuan \surname{Yao}}
\email{MX120170257@yzu.edu.cn}
\affiliation{Center for Gravitation and Cosmology, College of Physical Science and Technology, Yangzhou University, \\180 Siwangting Road, Yangzhou City, Jiangsu Province  225002, China}

\begin{abstract}
Hiscock and Weems showed that under Hawking evaporation, an isolated asymptotically flat Reissner-Nordstr\"om (RN) black hole evolves in a surprising manner:  if it starts with a relatively small value of charge-to-mass ratio $Q/M$, then said value will temporarily \emph{increase} along its evolutionary path, before finally decreases towards zero. This contrasts with highly charged ones that simply radiate away its charge steadily. The combination of these two effects is the cosmic censor at work:
there exists an attractor that flows towards the Schwazschild limit, which ensures that extremality -- and hence naked singularity -- can never be reached under Hawking evaporation. We apply the scheme of Hiscock and Weems to model the evaporation of an asymptotically flat dilatonic charge black hole known as the Garfinkle-Horowitz-Strominger (GHS) black hole. 
We found that upholding the cosmic censorship requires us to modify the charged particle production rate, which remarkably agrees with the expression obtained independently via direct computation of charged particle production rate on curved spacetime background. This not only strengthens the case for cosmic censorship, but also provides an example in which cosmic censorship can be a useful principle to deduce other physics.
We also found that the attractor behavior is not necessarily related to the specific heat, contrary to the claim by Hiscock and Weems.
\end{abstract}

\maketitle

\section{Introduction: Cosmic Censorship and Hawking Evaporation}\label{1}

Singularities are common occurrence in classical general relativity: the singularity theorem \cite{1,1-2,2,3,1410.5226,9606016} proves that they necessarily form under gravitational collapse during black hole formation. 
Since astrophysical observations have shown that the Universe is full of black holes, singularities are indeed widespread. If singularities can become naked under generic physical processes, then we lose classical predictability: given an initial data on a hypersurface, we cannot evolve the system  to determine the future uniquely, without knowing what boundary conditions to prescribe to the naked singularities. Therefore Penrose proposed the cosmic censorship conjecture, which essentially states that all black hole singularities would have to be concealed behind a horizon \cite{1-2}. More precisely, this is the \emph{weak} cosmic censorship conjecture, or in the famous words of Hawking, ``\emph{Nature abhors a naked singularity}''. 

The \emph{strong} cosmic censorship, on the other hand, requires that ``physically reasonable'' solutions to the Einstein field equations should be globally hyperbolic, i.e. the future is \emph{uniquely} predictable via the evolution equations once the initial and boundary conditions are given. Reissner-Nordstr\"om and Kerr black hole spacetimes do not satisfy global hyperbolicity due to the presence of a Cauchy horizon, beyond which the timelike singularity is ``naked'' to the interior observer. The strong cosmic censorship is generally believed to restore predictability by turning the Cauchy horizon into a singularity (generically spacelike) via the so-called ``mass inflation'' mechanism (divergence of energy density due to the blueshift of infalling light on the Cauchy horizon), though see however, \cite{1309.0224}. So, are naked singularities necessarily verboten? 
15000
The answer appears to be no: after some 50 years of research, it is now appreciated that there exist counter-examples to the cosmic censorship conjecture (see also \cite{0702116}). Perhaps the most well-known example is the finite time pinch-off of higher dimensional black objects \cite{1006.5960,1512.04532,1702.01755} due to the Gregory-Laflamme instability \cite{9301052, 9404071, 1906.10696} (for a review, see \cite{1107.5821}). Black hole collisions in higher dimensions also result in naked singularities \cite{1812.05017}. More recently, BTZ (Ba\~{n}ados-Teitelboim-Zanelli) black holes in 3-dimensions are also shown to violate the strong cosmic censorship \cite{1906.08265}. These are, however, not directly relevant to our actual, 4-dimensional Universe. 

The status of the cosmic censorship as far as physically relevant spacetimes are concerned, is currently being hotly debated. Recently various authors have shown that strong cosmic censorship may be violated for charged, near-extremal Reissner-Nordstr\"om black holes in an asymptotically \emph{de Sitter} spacetime (essentially by using cosmological redshift to weaken the blueshift at the Cauchy horizon), though the status is still somewhat unclear \cite{1711.10502, 1808.03631, 1811.08538, 1902.01865}. Even in vacuum asymptotically flat Kerr spacetimes (not necessarily near extremal), the work of Dafermos and Luk \cite{1710.01722} -- assuming that the exterior Kerr spacetime is indeed dynamically stable -- rigorously showed that the Cauchy horizon is not as singular as previously expected from mass inflation mechanism. Instead it forms a ``weak null singularity''\footnote{Intuitively, the singularity is weak enough such that an observer could conceivably pass through it. Technically, the singularity is weak enough such that the metric in an appropriate coordinate system is continuous up to the boundary. For details, see \cite{1311.4970}.} such that the metric can be extended beyond it, so strong cosmic censorship is violated in the strict sense. Nevertheless it is still not clear how one could follow the evolution beyond the weak null singularity as it cannot be made sense of even as a weak solution to the Einstein field equations. 

Remarkably, there is evidence to suggest that the cosmic censorship conjecture is related to the \emph{weak gravity conjecture} \cite{0601001}, which states that there should be at least one charged particle whose charge is bigger than its mass in Planck units (colloquially, gravity is the weakest force). Thus, in the presence of sufficiently charged particles, cosmic censorship can be saved (at least in anti-de Sitter spacetimes \cite{1709.07880, 1901.11096}). The situation in asymptotically de Sitter spacetimes remains unclear, as the presence of charged scalars and charged fermions do not seem to completely protect the cosmic censorship \cite{1808.03635, 1808.04832, 1811.10629} -- though it helps to some extent -- despite some earlier evidences for such a claim \cite{1801.07261, 1803.05443}. The situation in the presence of a massless Dirac field were discussed in \cite{1810.12128, 1905.06675}. 
It is clear from these huge literature and recent works that cosmic censorship is nowadays a very active area of research, with a lot of physics remain to be uncovered.

One might take a step back and ask whether we \emph{need} to worry about violation of the cosmic censorship, since singularities are likely to be 	``cured'' in a fully working theory of quantum gravity \cite{1605.06078, 1902.01583}, or otherwise prevented from ever arising since the black holes themselves could be de-stabilized by stringy effects \cite{1906.01169}. 
Although it is almost always taken as granted that singularity will be resolved by quantum gravity, but since we do not yet have such a theory, we do not know for sure. 
On page 5 of \cite{Hawking}, Hawking's point of view is that if the singularity is indeed
smeared out by quantum correction, it would be ``boring'' since gravity would then be ``just like any other field''. Hawking's point was that, notwithstanding a graviton description, gravity should be distinctively different since it is not just a player on a spacetime background; it is both a player
and the evolving stage. 

Indeed, there are genuine concerns about whether quantum gravity can solve this issue.
Sometimes, resolving a singularity might itself create other problems. For example, a Schwarzschild-AdS metric with negative mass corresponds to a naked timelike singularity. If this singularity is resolved, there would be states of arbitrarily negative energy, which would then contradict holography, which requires that the corresponding field theory Hamiltonian be bounded from below \cite{9503062,0410040}. Note that holography is really a correspondence between quantum gravity in the AdS bulk to field theory on its boundary. Although in practice one usually uses classical black hole solution, as an effective field theory the ``true'' action in the bulk has higher curvature terms, which would become important when discussing singularities. 
The negative energy solutions that remain after ``curing'' the singularity therefore do not contradict the AdS version of the positive energy theorem because the higher curvature terms effectively violate the local energy condition required by this theorem \cite{9503062}. 

This simply means that we do not yet understand issues involving singularities in string theory, thus one must consider the possibility that some form of singularities might persist even in quantum gravity. Therefore, given a specific process, e.g. in an attempt to either over-spin or over-charge a black hole, it is thus better if one could show that cosmic censorship cannot be violated, already at the classical level and/or semi-classical level.  

It is therefore a relief when Hiscock and Weems \cite{HW}, in 1990, investigated the evolution of an evaporating asymptotically flat Reissner-Nordstr\"om (RN) black hole and found that cosmic censorship is not violated (although they did not emphasize on cosmic censorship in that work). The details are, however, rather subtle. If the charge to mass ratio $Q/M$ is relatively small, then RN black holes first evolve towards extremality. This is not so surprising because the emission of massive particles is strongly suppressed in the Hawking radiation until the black hole gets sufficiently hot, and since there is no massless charged particle\footnote{Cosmic censorship conjecture in the presence of hypothetical massless charged particles was discussed in \cite{1709.05081}.} (at least in the standard model of particle physics).	This means that black hole tends to lose mass more than charge, and so $Q/M$ should increase. One might suspect that this would continue until the black hole becomes extremal (see, e.g., \cite{1710.02725, 1710.03783}). 

However, explicit detailed modeling by Hiscock and Weems demonstrated  that this does not continue indefinitely (see also \cite{page,gibbons}). Instead,  discharge eventually becomes  more efficient and $Q/M$ decreases. If $Q/M$ is relatively large from the start, then the black hole simply discharges and evolves steadily to decrease $Q/M$. These two different scenarios are referred to as the ``mass dissipation zone'' and ``charge dissipation zone'', respectively \cite{HW}.  Regardless, the final fate of an evaporating RN black hole is to lose its charge and follows an ``attractor'' (see Sec.(\ref{2}) for details) to approach a Schwarzschild configuration. It is worth noting that under such an evolution, starting from the mass dissipation zone, $Q/M$ can approach the extremal value, but it eventually turns over and so extremality cannot be reached, and the cosmic censorship is safe. 

In this work, we will also point out that the existence of attractor solution is not necessarily related to the specific heat of black holes; though Hiscock and Weems argue that the change of sign in the specific heat gives rise to the Reissner-Nordstr\"om attractor.\newline

\emph{A natural question to ask is whether such an attractor behavior is universal in some sense under Hawking evaporation}. By attractor behavior we do not necessarily mean that it attracts all evolutions toward Schwarzschild end state, but rather just pulls all evolutions away from ever reaching extremality. For example, an asymptotically flat Kerr black hole that only emits massless scalar particle would either spin up or spin down depending on its initial $a/M$ ratio towards $a/M\approx 0.555$, where $a=J/(Mc)$ is the usual angular momentum parameter, though adding higher spin fields tend to bring its final state to a Schwarzschild one unless there are sufficiently many species of massless scalars \cite{9710013, 9801044}. See also \cite{1508.06685}. It is therefore interesting to check more black hole spacetimes, including those beyond general relativity.

In this work we investigate this issue by employing the Hiscock-Weems technique to study the Hawking evaporation of a dilatonic charge black hole, known as the (Gibbons-Maeda-)Garfinkle-Horowitz-Strominger (GHS) black hole \cite{ghs,g,gm}, which is a well-motivated solution obtained from a low energy limit of string theory (see Sec.(\ref{2})), which remains physically viable. For some observational aspects of Einstein-Maxwell-dilaton black holes, see e.g., \cite{1706.06519, 1706.09875,1804.05812}. 

It should be emphasized that for GHS black holes, the extremal case is \emph{not} a zero temperature state with nonzero entropy, such as an extremal RN black hole\footnote{It has been suggested that for an extremal RN black hole, its horizon is effectively singular, i.e. a null singularity spacetime boundary \cite{1005.2999}. Therefore, perhaps extremal RN black holes do behave like extremal GHS one, and thus by studying GHS black holes, one may gain more insights into extremal RN black holes as well.}. It is also not an extremal vanishing horizon (EVH) spacetime, such as a 5-dimensional extremal Kerr black hole with one angular parameter set to zero, which has both zero temperature and zero entropy (the horizon essentially coincides with the singularity) \cite{1906.01169}. Instead, an extremal GHS black hole is a bizarre configuration that has zero size (with horizon coincides with a null singularity) but nonzero temperature\footnote{One might think that if a black hole has zero size, it is as good as having \emph{no} horizon, and should therefore have zero temperature. However, this is not the case. Prescribing extremal ``hole'' zero temperature -- or indeed any other temperature other than the non-extremal form $T=\hbar c^3/(8\pi G k_B M)$ -- would result in a divergent stress-energy \cite{9607048v1}. In this case, \emph{zero is not the same as nothing.}}. In fact, its Hawking temperature, with all physical constants explicitly written out, takes the same form as a Schwarzschild black hole with the same mass: $T=\hbar c^3/(8\pi G k_B M)$, where $M$ is the final mass of the black hole, related to its final charge by the extremal condition $Q^2/(4\pi\epsilon_0 c^4)=2M^2$, where $\epsilon_0$ is the vacuum permittivity. As usual, $G$ denotes Newton's gravitational constant, $k_B$ the Boltzmann constant, $\hbar$ the Planck constant, and $c$ the speed of light. 
Note that the extremal charge is $\sqrt{2}$ times greater than the RN case. Intuitively, this larger extremality value can be understood as due to an ``extra gravity'' contributed by the scalar field \cite{1901.11096}.

For EVH spacetimes, arguably the third law of black hole thermodynamics might apply, and one never really achieves the extremal configuration, which would otherwise be problematic because it is effectively a naked singularity: though the horizon coincides with the singularity, it is unstable \cite{0308056}. However, it is far from obvious that the usual third law of black hole mechanics applies to the extremal GHS ``black hole'' due to its nonzero temperature (in fact our numerical calculation in Sec.(\ref{3-2}) shows that these black holes can indeed become extremal in finite time, within numerical accuracy). Thus, there is a genuine concern that Hawking evaporation would bring a \emph{generic} GHS black hole to extremal state, thus once perturbation is taken into account\footnote{It was shown that an extremal charged dilaton black hole could be destroyed by injecting a test particle with a specific energy. Nevertheless the censorship is well protected if backreaction or self-force is included \cite{1803.07916}. This does not exclude the possibility that other types of perturbations could render the singularity naked. In fact, Ref.\cite{1803.07916} assumed the black hole to be static, whereas an earlier study with dynamical solution found that the black hole can be overcharged by bombarding it with a stream of electrically charged null fluid, resulting in a naked singularity \cite{1512.08550}. See, however, \cite{1703.07414} and \cite{1812.06966}.}, becomes a naked timelike singularity and thus potentially violates the cosmic censorship within the scope of low energy, effective string theory\footnote{For previous study of singularities in such regimes, see, e.g., \cite{9804039,1703.07414,1512.08550}.}. 
Earlier study has indeed suggested that dilatonic charged black holes harbor the danger of becoming extremal \cite{9508029v1}. In this work we provide additional evidence to support this claim.

In fact, even if we do not consider any perturbation to the extremal configuration, one can show that the Kretschmann scalar diverges in the extremal limit, and so the spirit of the cosmic censorship is already violated. In addition, in our work, since the black hole mass remains large in this limit, the violation would be much worse than the ``mild'' violation of the Gregory-Laflamme type, in which the singularity occurs at the ``necks'' with a very small mass density \cite{1812.05017}.

In a previous study by the authors, a ``complementary third law'' for black hole thermodynamics was found \cite{comp}. It states that for a large class of black holes whose end state has finite nonzero temperature but zero size, the evolution time is necessarily infinite (as opposed to the usual third law, which states that the evolution towards zero temperature takes an infinite time). Although the result therein was only proved for neutral black holes, GHS black holes were shown to satisfy the complementary third law if the charge is held fixed while allowing mass to evaporate. It was then conjectured that since the temperature does not depend on the charge, as opposed to the RN case, allowing the charge to evaporate should not affect the result by much. Our current work shows that this conjecture is not correct, just like the naive expectation that RN black holes could evolve towards extremality. We will come back to this point in the Discussion, Sec.(\ref{4}).

The main result we obtained in this work is that GHS black hole \emph{cannot} attain extremality under Hawking evaporation, and so the cosmic censorship is safe under this context. \emph{This, however, is only so because remarkably the dilaton coupling to the Maxwell field is such that charged particle production rate is modified in exactly the right way needed to ensure extremality is avoided}. Colloquially, one might even say that there is a conspiracy to uphold censorship.

The organization of this work is as follows: in Sec.(\ref{2}), we will first review the original work of Hiscock and Weems \cite{HW}, which investigated the evolution of RN black holes under Hawking evaporation. In Sec.(\ref{3}) we will apply the same technique to the study of GHS black holes, and further discuss how cosmic censorship remains valid in Sec.(\ref{3-2}) once dilaton induced corretion is taken into account. We conclude with some discussions in Sec.(\ref{4}), and raise some questions for future studies.

Henceforth, we will follow the convention of Hiscock and Weems, employing the ``relativistic units'' \cite{werner} in which Newton's
gravitational constant $G$, the speed of light $c$, and the Boltzmann constant $k_B$ are set to unity. The vacuum permittivity is set to satisfy $4\pi \epsilon_0=1$. However, the reduced Planck constant $\hbar$ is \emph{not} set to 1.
This differs from the Planck units. Thus, in this work we have $\hbar=\hbar G/c^3 \approx 2.61 \times 10^{-66} \text{cm}^2$, while e.g., a solar mass is  $M_\odot = M_\odot G/c^2 \approx 1.5 \text{km}$. Mass and charge both have the dimension of length. In all plots, the units for mass and charge are centimeters.
Temperature has the dimension of length in this unit, instead of dimension of inverse length in the more oftenly used natural units.

\section{Hiscock and Weems Model for Evaporating Reissner-Nordstr\"om Black Holes}\label{2}

In this section we will review the models and results of Hiscock and Weems \cite{HW}, so that our work  is self-contained, especially when we need to compare our GHS black holes results with the RN counterparts later.

Hiscock and Weems \cite{HW} considered an isolated\footnote{This is because a black hole surrounded by realistic astrophysical medium will discharge quickly. A quick estimate given in \cite{HW} is that for the black hole to avoid rapid discharge by accretion, the Newtonian gravitational attraction must be larger than the electromagnetic Coulomb attraction and repulsion for the lightest charged particle pair, namely electron and positron. Thus $Mm/r^2 \gg Qe/r^2$, so that in the presence of accretion medium, discharge will quickly bring the charge-to-mass ratio down to $Q/M \ll m/e \approx 10^{-21}$.} asymptotically flat Reissner-Nordstr\"om black hole, with metric tensor given by
\begin{flalign}
g[\text{RN}]=&-\left(1-\frac{2M}{r}+\frac{Q^2}{r^2}\right)\d t^2\\ \notag & + \left(1-\frac{2M}{r}+\frac{Q^2}{r^2}\right)^{-1}\d r^2 + r^2\d\Omega^2_{S^2},
\end{flalign}
where $\d\Omega^2_{S^2}$ is the round metric of a 2-sphere. 
Without loss of generality, the charge $Q$ is assumed to be positive. The event horizon is located at $r_+=M+\sqrt{M^2-Q^2}$. In modeling the Hawking emission, all scattering is neglected. This is a good approximation since the created particles have the same sign charge as the black hole, so the black hole will experience an enormous radial repulsive force.

Then Hiscock and Weems considered the black hole to be sufficiently large. This is for the following reason: their idea is to model charge loss using Schwinger formula, while only allows massless particle to be governed by the Stefan-Boltzmann equation. This means that the black hole should be cold enough so that production of massive charged particles are negligible. Since the temperature is inversely proportional to the mass\footnote{This is obvious for a Schwarzschild black hole, but it is also true for RN black hole, whose temperature is $T=\frac{\hbar(M^2-Q^2)}{2\pi \left(M+\sqrt{M^2-Q^2}\right)^2}$, since it is easily seen that, due to $Q \leqslant M$, 
\begin{equation}\frac{\partial T}{\partial M}=-\frac{1}{8\pi}\left[\frac{M^3-3MQ^2+(M^2-Q^2)\sqrt{M^2-Q^2}}{M^2 (M+\sqrt{M^2-Q^2})^3} \right] <0. \notag
\end{equation}
},
this means we want a sufficiently massive black hole.
Indeed, as shown by Gibbons \cite{gibbons}, black holes with radius larger than the reduced Compton wavelength of the electron pair-produce electron-positron pairs electromagnetically instead of gravitationally. 
Note that as emphasized by Hiscock and Weems \cite{HW}, although the production of charged particles is treated separately from the thermal Hawking flux of neutral particles, this is just part of the modeling valid within the appropriate regime; in actual fact they are all part of the Hawking radiation. 

Indeed, although we usually say that extremal black holes with zero temperature do not radiate, this is not true -- they can radiate via the charge channel. Hawking calculated that the number of particles of the $j$-th species
with charge $q$ emitted in a wave mode labeled by frequency $\omega$, spheroidal harmonic $l$, and
helicity $p$ is given by (if we ignore angular momentum of emitted particles and rotation of
the hole) \cite{Hawking2}, 
\begin{equation}\label{Hawking}
\left\langle N_{j\omega l p} \right\rangle = \frac{ \Gamma_{j\omega l p}}{\exp\left[(\omega-e\Phi)/T\right]\pm 1}, 
\end{equation}
where $T$ is the temperature of the black hole. The plus sign in the denominator corresponds to fermion, while minus sign corresponds to boson. Here $ \Gamma_{j\omega l p}$ denotes the absorption probability for an incoming wave of the specific mode. For sufficiently strong electric field, even if $T=0$, particle number need not be zero. Whether this non-thermal part of particle emission should be called ``Hawking radiation'' is somewhat a semantic issue, in fact in some sense the Schwinger mechanism and the Hawking radiation are generically indistinguishable for near extremal black holes \cite{1202.3224}. Readers who prefer to interpret only the thermal part as Hawking radiation, as is commonly done in the literature, are free to do so. Then Hiscock-Weems model should be thought of as black hole evaporation under both Hawking emission and Schwinger pair production. In this work however, we follow Hiscock and Weems' convention and refer to everything as Hawking radiation.

In Hiscock-Weems method, for a black hole to be ``sufficiently massive'' means $M \gg \hbar/m_e \approx 10^{-10}\text{cm}\approx 10^8\text{g}$ where $m_e$ is the electron mass\footnote{Though, since we consider $Q>0$, in the following analysis we will actually be dealing with the positron.}. In this mass range the flat spacetime quantum electrodynamics is a good approximation \cite{HW, gibbons}.

Next, in order to be able to ignore the higher order terms in the Schwinger formula, 
\begin{equation}
\Gamma = \frac{e^2}{4\pi^3\hbar^2}\frac{Q^2}{r^4}\exp\left(\frac{-\pi m^2 r^2}{\hbar e Q}\right)\times\left[1+{O}\left(\frac{e^3Q}{m^2r^2}\right) + \cdots\right], \label{Schwinger}
\end{equation}
where $\Gamma$ is the rate of electron-positron pair creation per unit 4-volume, Hiscock and Weems considered the weak-field regime, which requires $e^3Q/m^2r^2 \ll 1$ for all $r\geqslant r_+$. This requires the mass range to be $M \gg e^3/m^2 \approx 6 \times 10^8\text{cm} \approx 4 \times 10^3 M_\odot$. 

Finally, in order to use certain asymptotic series expansion, which we will see the details below, Hiscock and Weems also needed to impose $M \gg Q_o:=\hbar e/\pi m^2 \approx 6.47 \times 10^{10}\text{cm}$. 

By combining all the above assumptions and their valid regimes, we are essentially dealing with mass range $M \gg Q_0$, which is of the order $10^{10}\text{cm}$ or $10^5$ solar masses.

Now let us look at the explicit equations. The charge loss can be obtained from the Schwinger formula:
\begin{equation}\label{dQdt0}
\frac{\d Q}{\d t} \approx -\frac{e^2}{\pi^2\hbar^2} \int_{r_+}^\infty \frac{Q^2}{r^2}\exp\left(-\frac{r^2}{Q_0Q}\right) \d r \times e.
\end{equation}
The integral can be expressed in terms of the complementary error function, which can be expanded as
\begin{equation}
\text{erfc}(x)=\frac{e^{-x^2}}{x\sqrt{\pi}}\left[1+\sum_{n=1}^{\infty}(-1)^n\frac{1\cdot3\cdot5\cdots (2n-1)}{(2x^2)^n}\right],
\end{equation}
with $x \gg 1$ being replaced by $r/\sqrt{Q_0Q}$.
This is where we require $M \gg Q_0$. The end result is the charge loss ODE 
\begin{equation}\label{dQdt}
\frac{\d Q}{\d t} \approx -\frac{e^4}{2\pi^3\hbar m^2}\frac{Q^3}{r_+^3}\exp\left(-\frac{r_+^2}{Q_0Q}\right).
\end{equation}
We remark that
$Q_0$ is essentially the inverse of the Schwinger critical field: $E_c:={m^2 c^3}/{e\hbar} = 1.312 \times 10^{16}~\text{V/cm}$. If the electric field is stronger than $E_c$, then charged particles are created from the vacuum at an exponential rate. In the model of Hiscock and Weems, this never happens: so although there are charged particles being produced, the rate is suppressed \cite{gibbons, 1404.5215}. To see this, simply note that the series approximation requires $r_+^2 \gg QQ_0$, i.e. $Q/r_+^2 \ll E_c$.
Eq.(\ref{dQdt}) can also be obtained via the Wentzel-Kramers-Brillouin (WKB) approximation \cite{1305.2564}, a point we will return to in Sec.(\ref{3-2}).

Since we are dealing with cold black holes, we may assume that thermal part of the radiation only contains massless particles. So the mass loss rate is
\begin{equation}\label{dMdt}
\frac{\d M}{\d t} = -a \alpha \sigma T^4 + \frac{Q}{r_+}\frac{\d Q}{\d t},
\end{equation}
where the second term is from the first law of black hole thermodynamics, due to (suppresed) massive particle emission, which dominates at either low $T$ or, equivalently, at high $Q$. Here $a=\pi^2/(15\hbar^3)$ is the radiation constant, i.e. $4/c$ times the Stefan-Boltzmann constant\footnote{Hiscock and Weems simply refer to $a$ as the Stefan-Boltzmann constant.}, while $\sigma$ is the geometric optics cross section,
\begin{equation}
\sigma[\text{RN}]=\frac{\pi}{8}\frac{(3M+\sqrt{9M^2-8Q^2})^4}{(3M^2-2Q^2+M\sqrt{9M^2-8Q^2})},
\end{equation}
whose radius corresponds to the impact parameter of massless particle from infinity that ends up on the photon orbit (of course here we want the particle to escape to infinity; the condition is the same).  The reason is that only particles that have enough energy can escape the effective potential barrier. Due to this potential barrier, the Hawking radiation received at asymptotic region is not a black body radiation. The deviation from black body is governed by the constant $\alpha$, the greybody factor, which is different for different species of particles (see \cite{HW} for more discussions). For simplicity we may set $\alpha$ to be unity. Compared to the long lifetime of the black hole, its effect is negligible for our purpose.

Eq.(\ref{dQdt}) and Eq.(\ref{dMdt}) therefore form a coupled set of ordinary differential equations that govern the evolution of RN black holes under Hawking evaporation. In general this system can only be solved numerically. In their paper, Hiscock and Weems discussed in details the evaporation history of RN black holes given different initial conditions. The most important results are summarized by their Fig.(2). We repeated their calculations and reproduced it in Fig.(\ref{RN-evolution}) below. It shows the evolutionary paths of evaporating RN black holes.

Let us emphasize again the main features in the plot. There are two distinct regions: the charge dissipation regime and the mass dissipation regime. 
The charge dissipation regime corresponds to highly charged initial conditions, the black holes simply radiate off its charge steadily and evolve towards Schwarzschild final state. What happens afterward depends on whether the Stefan-Boltzmann law continues to hold indefinitely, in which the black hole completely evaporates in finite time, or if new physics comes into play at sufficiently high energy, which could halt the evaporation and ends with a ``remnant'' \cite{aharonov, 1412.8366}. 

On the other hand, if a GHS black hole starts with a low charge-to-mass ratio, then it is in the mass dissipation regime: its Hawking radiation is dominated by massless neutral particles, so the black hole loses mass but retains much of its charge. Therefore initially the  charge-to-mass ratio increases steadily, until the evolutionary path hits the attractor, and then the evolution turns around and approaches the Schwarzschild state. Although $Q/M$ can approach the extremal value of unity, the evolution always turns around before extremality is reached, therefore upholding the cosmic censorship. 

\begin{figure}[!h]
\centering
\includegraphics[width=3.4in]{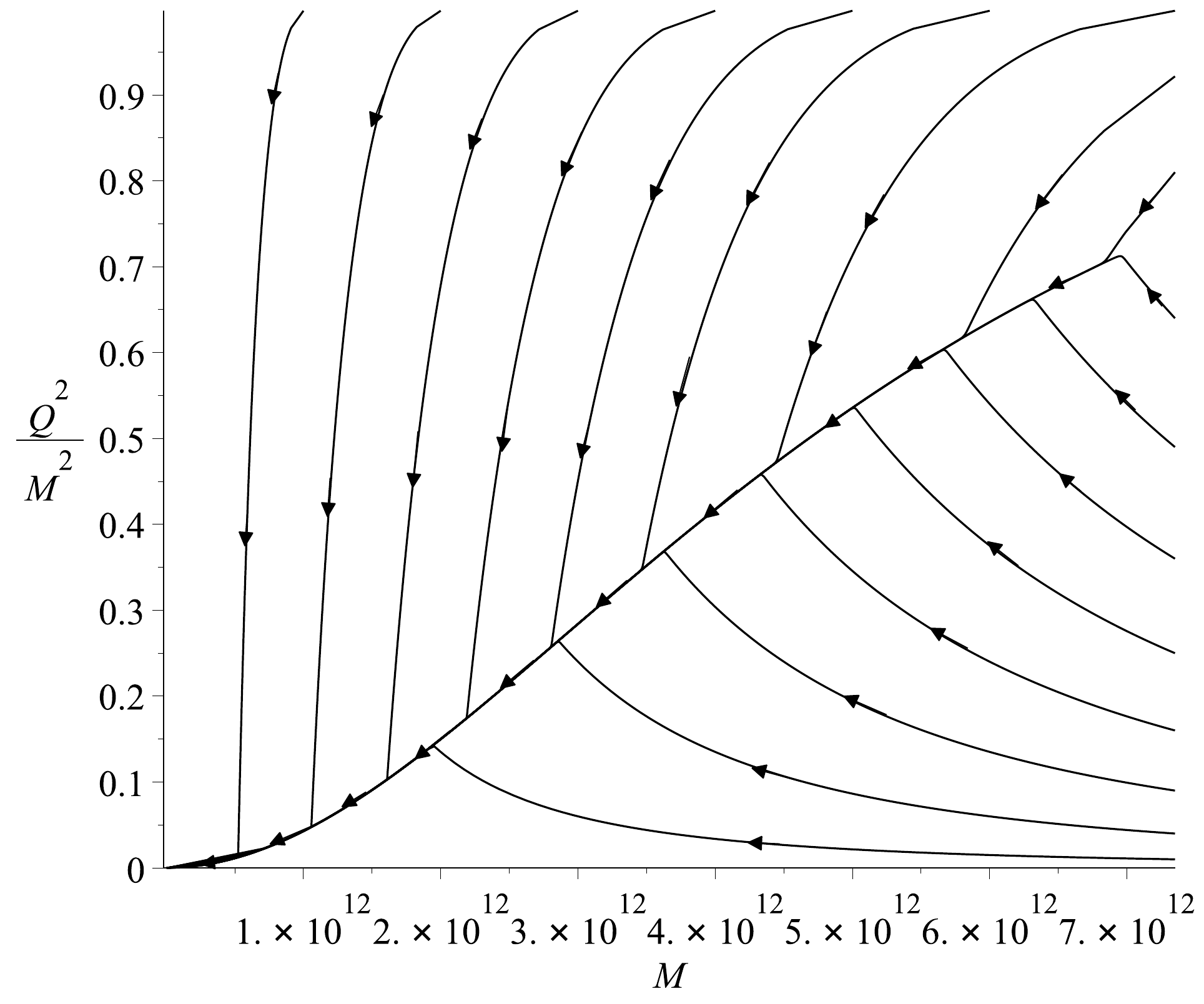}
\caption{The evolution of asymptotically flat Reissner-Nordstr\"om black holes under Hawking evaporation with Hiscock-Weems model. Each diagram in this point corresponds to a specific mass and charge. Given an initial condition, it will evolve by following the arrows. A black hole with sufficiently large charge-to-mass ratio simply discharges towards the Schwarzschild limit: this is the ``charge dissipation regime''. On the other hand, a black hole starting with relatively small charge-to-mass ratio will lose its mass faster than losing charge, and thus $Q/M$ increases: this is the ``mass dissipation regime'' . These curves eventually hit the attractor and flow towards Schwarzschild limit. The model breaks down for sufficiently small black holes ($M  \lesssim Q_0$). Black holes that are too small will discharge any geometrically interesting charge in a time shorter than the characteristic time associated with the geometry of the black hole \cite{HW}.} \label{RN-evolution}
\end{figure}

Hiscock and Weems argued that the existence of the attractor is due to the fact that the specific heat of RN black hole changes sign.
While they did show that the attractor might be related to the specific heat, as we will argue below with the GHS case, this is not the underlying reason that \emph{causes} the attractor to exist, since in the GHS case the specific heat simply does not change sign.

\section{The Evolution of Hawking Evaporating GHS Black Holes}\label{3}

We shall now move on to discuss the evaporation of GHS black holes under the Hiscock-Weems model, starting from a short review on the GHS black hole spacetime. 

The 4-dimensional low-energy Lagrangian obtained from heterotic string theory is\footnote{Note that the action is similar to that of the 4-dimensional effective action reduced from 5-dimensional Kaluza-Klein theory \cite{1107.5563}:
\begin{equation}
S[\text{KK}]= \frac{1}{16\pi}\int \text{d}^4 x \sqrt{-g}\left[R - 2(\nabla \varphi)^2 - e^{-2\sqrt{3}\varphi}F^2\right],
\end{equation}
where an additional $\sqrt{3}$ appears in the exponential term. The Kaluza-Klein dimensional reduction is however different from the string theory approach, see further discussions in \cite{9210119}.}
\begin{equation}\label{action}
S[\text{EDM}]= \frac{1}{16\pi}\int \text{d}^4 x \sqrt{-g}\left[R - 2(\nabla \varphi)^2 - e^{-2\varphi}F^2\right],
\end{equation}
where $F^2:=F_{\mu\nu}F^{\mu\nu}$, with $F_{\mu\nu}$ being the (components of) Maxwell field, $F=(Q/r^2) \d t \wedge \d r$, associated with a U(1) subgroup of $E_8 \times E_8$ or $\text{Spin}(32)/\Bbb{Z}_2$, and $\varphi$ is the dilaton, a scalar field coupled to the Maxwell field. The remaining gauge fields and antisymmetric 3-form field $H_{\mu\nu\rho}$ (related to the axion) are set to zero \cite{ghs}. 

This is an example of Einstein-Maxwell-Dilaton theory.
In general the (massless) dilaton field could have a finite value $\varphi_0$ at spatial infinity, which we have set to be zero in this work. 

The GHS black hole solution takes the deceptively simple form\footnote{This is of course a static solution, whereas here we are considering a dynamical evolution. A more rigorous approach would be to use a dynamical solution, a Vaidya-type generalization of the GHS metric \cite{1703.07414}. Nevertheless, Hawking evaporation takes a very long time, so quasi-static approximation is good enough for our purpose, especially in the large $M$ regime we are investigating.}
\begin{flalign}\label{GHSg}
g[\text{GHS}]=&-\left(1-\frac{2M}{r}\right)\d t^2 + \left(1-\frac{2M}{r}\right)^{-1}\d r^2 \notag \\ &+ r\left(r-\frac{Q^2}{M}\right)\d \Omega^2_{S^2}.
\end{flalign}
The $r$-$t$ ``plane'' takes the Schwarzschild form and is independent of the charge $Q$. Thus the (only) horizon is \emph{always} located at $r_+=2M$. In the extremal case, $r_+=2M$ and $Q^2=2M^2$, so the radius of the $S^2$ part of the metric is equal to zero. That is to say, the extremal ``black hole'' has zero size; it is a null singularity. This has no contradiction with $r_+$ always remaining finite at $2M$, since $r$ is only a coordinate and not the areal radius. One could always switch to the areal radius $R:=[r(r-Q^2/M)]^{1/2}$, and then it is clear that the horizon shrinks to zero size for the extremal case. In either coordinate system, one could compute the Hawking temperature with the usual method of Wick-rotation by ensuring regularity at the Euclidean ``horizon''. This would indeed give the Schwarzschild form $T=\hbar/(8\pi M)$ for the temperature\footnote{In the GHS coordinates, as in Eq.(\ref{GHSg}), this is clear, since the Wick-rotation only concerns the $r$-$\tau$ ``plane'', where $\tau=it$ is the imaginary time. If we use the areal radius instead, the (Euclidean) metric would take the form $\d s^2=f(R)\d \tau^2 + g(R) \d r^2 + R^2 \d \Omega_{S^2}^2$, such that $fg^{-1}\neq 1$. With prime denoting differentiation with respect to $R$, one could show that the Hawking temperature is $T=\hbar\sqrt{f'g'}/{4\pi}$, which in this case gives the same result.}, independent of the charge $Q$. 

Note that 
due to the coupling with the Maxwell field, the dilaton is not an independent ``hair'' of the black hole. In fact, the dilaton is related to the Maxwell charge $Q$ by \cite{ghs, 9210119}
\begin{equation}
e^{2\varphi}=1-\frac{Q^2}{Mr},
\end{equation}
so $\varphi=0$ implies $Q=0$ and vice versa. That is, the dilaton does not exist independently of the electrical charge; due to the coupling between the dilaton and the Maxwell fields,  the GHS solution does \emph{not} recover the RN solution when $\varphi \to 0$. 

As explained before,
the second term in the mass loss equation Eq.(\ref{dMdt}) is simply from the first law of black hole thermodynamics. Since the dilaton is a secondary hair, the dilaton charge \cite{1806.10238} is a ``redundant parameter'' \cite{1612.09279}, which does not appear in the first law (see, however, related discussions \cite{1803.11317, 9607108}). It therefore suffices to consider the evolution of the primary hairs $M$ and $Q$ (see also \cite{9508029v1}). Thus,
the mass loss equation is the same as Eq.(\ref{dMdt}), except that the Hawking temperature is now $T=\hbar/(8\pi M)$ and the
geometric optic cross section needs to be replaced with \cite{1109.0254}
\begin{equation}
\sigma[\text{GHS}]=\frac{r_\text{ph}^2\left(r_\text{ph}-\frac{Q^2}{M}\right)}{r_\text{ph}-2M},
\end{equation}
where $r_\text{ph}$ is the GHS photon orbit \cite{0005050, 1109.0254}
\begin{equation}
r_\text{ph}=\frac{1}{4}\left[\frac{Q^2}{M} + 6M + \left(36M^2-20Q^2 + \frac{Q^4}{M^2}\right)^{\frac{1}{2}}\right].
\end{equation}
For consistency check, we note that these expressions reduce to the Schwarzschild case $\sigma[\text{Sch}]=27 M^2$ and $r_\text{ph}[\text{Sch}]=3M$ when $Q=0$.

To compare the results with the RN case investigated by Hiscock and Weems, we shall follow their scheme and work within the geometric optics cross section. Indeed, due to scattering at long wavelengths, the effective emission surface is smaller than the one given by the geomertic optics cross section. This is governed by the $\alpha$ greybody factor in Eq.(\ref{dMdt}), which we neglect in this work. Its inclusion tends to lengthen the lifetime of a black hole by a few order of magnitudes at most, and in our case we are not really concern about the exact lifetime, but rather how the charge-to-mass ratio evolves. 

\begin{figure}[!h]
\centering
\includegraphics[width=3.4in]{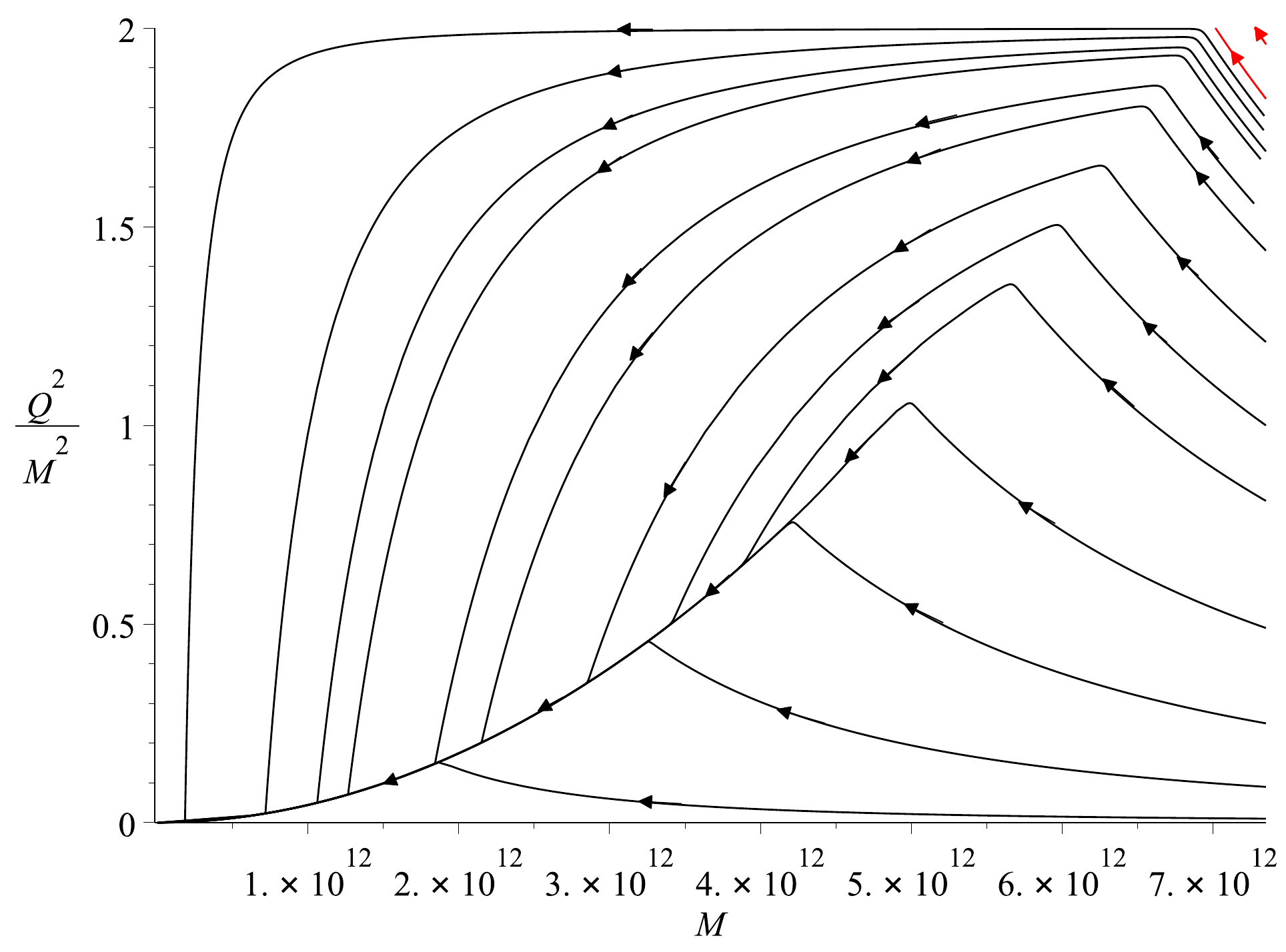}
\caption{The evolution of asymptotically flat GHS black holes under Hawking evaporation with Hiscock-Weems model, assuming $\d Q/\d t$ follows Eq.(\ref{dQdt}), shows that sufficiently massive black holes (in red) appear to evolve towards extremality. The extremality condition is $Q^2/M^2=2$.
\label{ghs-wrong-evolution}}
\end{figure}

\begin{figure}[!h]
\centering
\includegraphics[width=3.3in]{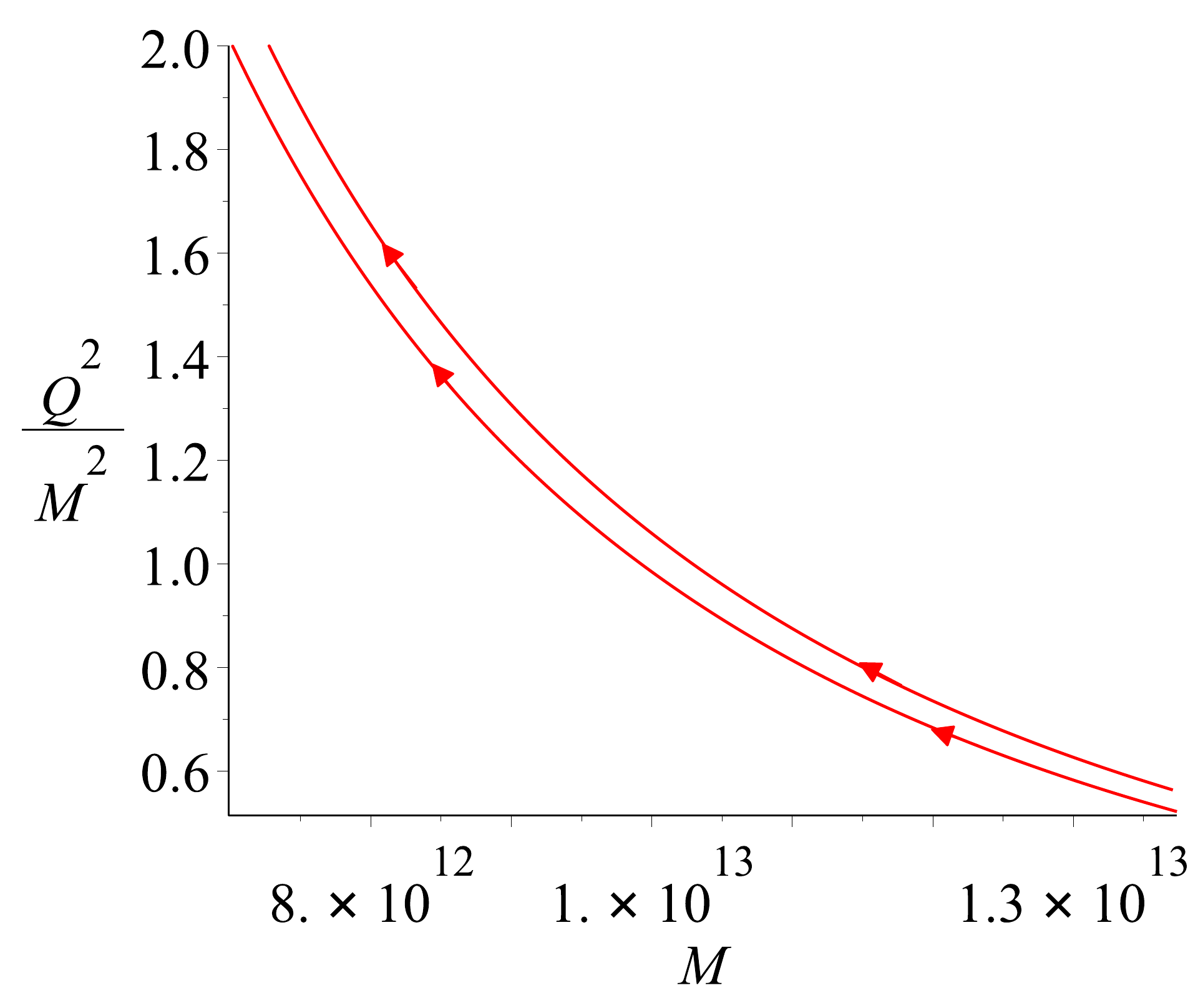}
\caption{Examples of GHS black holes that evolve towards extremality ($Q^2/M^2=2$) with Hiscock-Weems model. Parts of these curves are the same ones that appear in Fig.(\ref{ghs-wrong-evolution}).
\label{ghs-wrong-evolution-2}}
\end{figure}

If we naively use Eq.(\ref{dQdt}) for the charge loss rate and solve the coupled ODEs numerically, we find that although the features of Fig.(\ref{RN-evolution}) are largely retained over the same mass range, as shown in Fig.(\ref{ghs-wrong-evolution}), for sufficiently large mass, $(Q/M)^2$ appears to increase as the black hole evaporates, with no sign of turning around toward the Schwarzschild limit. Two such curves in Fig.(\ref{ghs-wrong-evolution}) are singled out in Fig.(\ref{ghs-wrong-evolution-2}) for a clearer view: indeed we see that starting at some low value of $Q/M$ a GHS black hole could evolve toward extremality in finite time. As discussed in the Introduction, Sec.(\ref{1}), this is in danger of violating cosmic censorship.

\section{Two Roads to Charged Particle Production Rate: Or The Conspiracy to Uphold Censorship}{\label{3-2}}

In the previous section we have shown that cosmic censorship can potentially be violated in GHS spacetimes, since extremality can be reached under Hawking evaporation. 
However, since the dilaton is exponentially coupled to Maxwell field in this theory, it is likely that charged pair production rate needs to be modified. 
In fact, as pointed out by Horowitz and Santos \cite{1901.11096} in a recent work, in which they investigated how cosmic censorship in dilaton theory is related to a revised weak gravity conjecture, the Maxwell field is smaller
(for the same source) when the dilaton is present. So, how might one derive the correct charge loss formula in the presence of dilaton?

In this work, we claim that imposing cosmic censorship allows us to \emph{derive} the dilaton induced correction to the charge loss rate, with only a mild, reasonable, assumption on the ansatz. We then show that this modification can be obtained independently via another consideration that has nothing to do with cosmic censorship, via direct consideration of wave scattering on the curved background. 

In other words, the cosmic censorship continues to be valid because the dilaton ``conspires'' with the Maxwell field in exactly the right way required to prevent extremality being reached by Hawking evaporating black holes.  

We first note that the reason $Q/M$ can increase is that charge loss is inefficient relative to mass loss. Therefore one could increase the efficiency by adding a correction term to the exponent of Eq.(\ref{dQdt}). Since $Q_0$ sets the physical scale, the simplest correction term would be of the form $\mathcal{C}{Q}/{Q_0}$ for some constant $\mathcal{C}>0$. That is, Eq.(\ref{dQdt}) becomes
\begin{equation}\label{dQdtnew}
\frac{\d Q}{\d t} \approx -\frac{e^4}{2\pi^3\hbar m^2}\frac{Q^3}{r_+^3}\exp\left(-\frac{r_+^2}{Q_0Q} + \mathcal{C}\frac{Q}{Q_0}\right).
\end{equation}
In order to fix $\mathcal{C}$, consider near an extremal GHS black hole, so that $Q=\sqrt{2} M - \varepsilon$, for some small $\varepsilon >0$. The exponential term can be expanded into series in $\varepsilon$:
\begin{flalign}
\exp\left(-\frac{r_+^2}{Q_0Q} + \mathcal{C}\frac{Q}{Q_0}\right) = &\exp\left(\frac{\sqrt{2}M (\mathcal{C}-2)}{Q_0}\right) \\ \notag &\times \left[1-\frac{2+\mathcal{C}}{Q_0}\varepsilon + O(\varepsilon^2)\right].
\end{flalign}
As long as $\mathcal{C} < 2$, the exponent is being suppressed by the leading order term. The suppression is greater for large $M$, this is why $Q/M$ for sufficiently large $M$ increases steadily to extremality. To phrase it in another way: for any fixed $\mathcal{C} < 2$, one could always choose $M$ sufficiently large so that there exist initial datum that lead to extremality under Hawking evaporation. 

For example, if we set $\mathcal{C}=1$, then both curves in Fig.(\ref{ghs-wrong-evolution-2}) do turn around toward the Schwarzschild limit, but this alone does \emph{not} guarantee other curves, which start with higher initial mass, to do the same. To be more concrete, we give an example: starting with mass $M(0)=7.35 \times 10^{13}$cm and $Q(0)=1.35 M(0)$, the black hole whose charge loss rate is given by Eq.(\ref{dQdtnew}), with $\mathcal{C}=1$, will reach extremality in finite time (albeit a long time, about $10^{93}$ years). This is depicted in Fig.(\ref{compare2}) below.

\emph{To guarantee that all curves turn around towards the Schwarzschild limit, we only need to set $\mathcal{C}=2$}. Then the leading order term is independent of the black hole mass. In fact the exponent becomes unity in the extremal case. The physical interpretation is that, as black holes evolve and come closer toward extremality, charged particle production rate is enhanced \emph{just enough} to turn the evolution around and send it towards the attractor, which then flows towards the Schwarzschild limit. The argument of the exponent never exceeds unity, so charged particles are never produced by a copious amount. 

\begin{figure}[!h]
\centering
\includegraphics[width=3.5in]{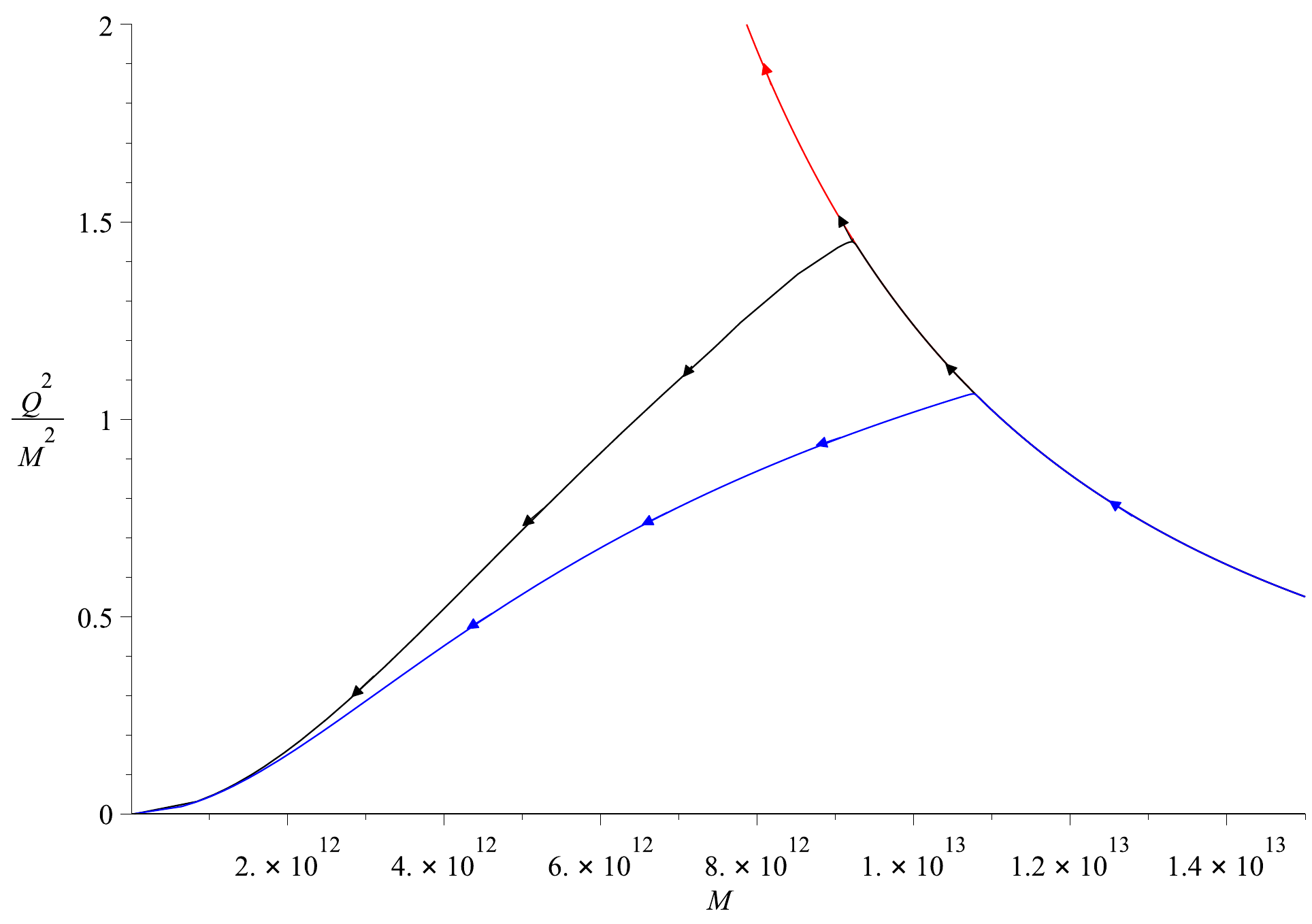}
\caption{GHS black hole evolution with initial conditions $M(0)=1.5\times 10^{13}$cm, $Q(0)=\sqrt{0.55} M(0)$. The curves, from top to bottom (red, black, blue), correspond to the choice $\mathcal{C}=0,1,2$, respectively, in the charge loss formula Eq.(\ref{dQdtnew}). Arrows denote the direction of evolution.}
\label{compare1}
\end{figure}

To illustrate our discussions thus far, in Fig.(\ref{compare1}), we show the evolution path of GHS black hole given different values of $\mathcal{C}$. Let us choose the initial conditions  $M(0)=1.5\times 10^{13}$cm, and $Q(0)=\sqrt{0.55} M(0)$. We note that for $\mathcal{C}=0$ (i.e. the original Schwinger formula), the black hole evolves towards extremality. With $\mathcal{C}=1$, it turns around and evolves towards Schwarzschild limit. The turn-around occurs earlier for $\mathcal{C}=2$.

 In Fig.(\ref{compare2}), we show that with a larger initial mass, $M(0)=7.35\times 10^{13}$ and $Q(0)=1.35 M(0)$, the black hole evolves towards extremality for $\mathcal{C}=0$. Setting $\mathcal{C}=1$ does not help in this case, the evolution is almost the same (the curves almost overlap). Setting $\mathcal{C}=2$, however, ensures that the curve turns around towards Schwarzschild limit.

\begin{figure}[!h]
\centering
\includegraphics[width=3.5in]{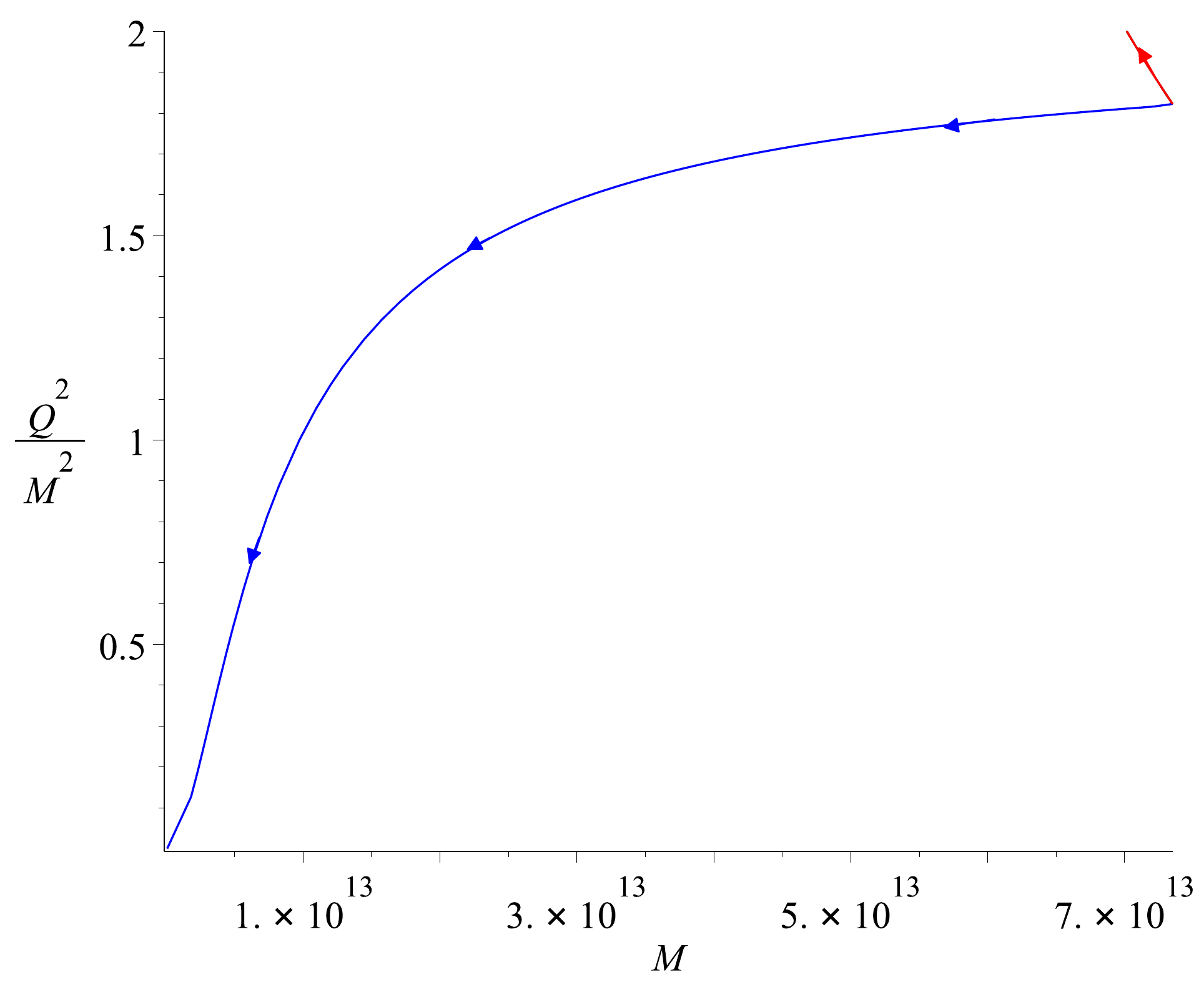}
\caption{GHS black hole evolution with initial conditions $M(0)=7.35\times 10^{13}$ and $Q(0)=1.35 M(0)$. The curves, from top to bottom (red, black, blue), correspond to the choice $\mathcal{C}=0,1,2$, respectively, in the charge loss formula Eq.(\ref{dQdtnew}). The top two curves essentially overlap in this plot. Arrows denote the direction of evolution.}
\label{compare2}
\end{figure}

\begin{figure}[!h]
\centering
\includegraphics[width=3.3in]{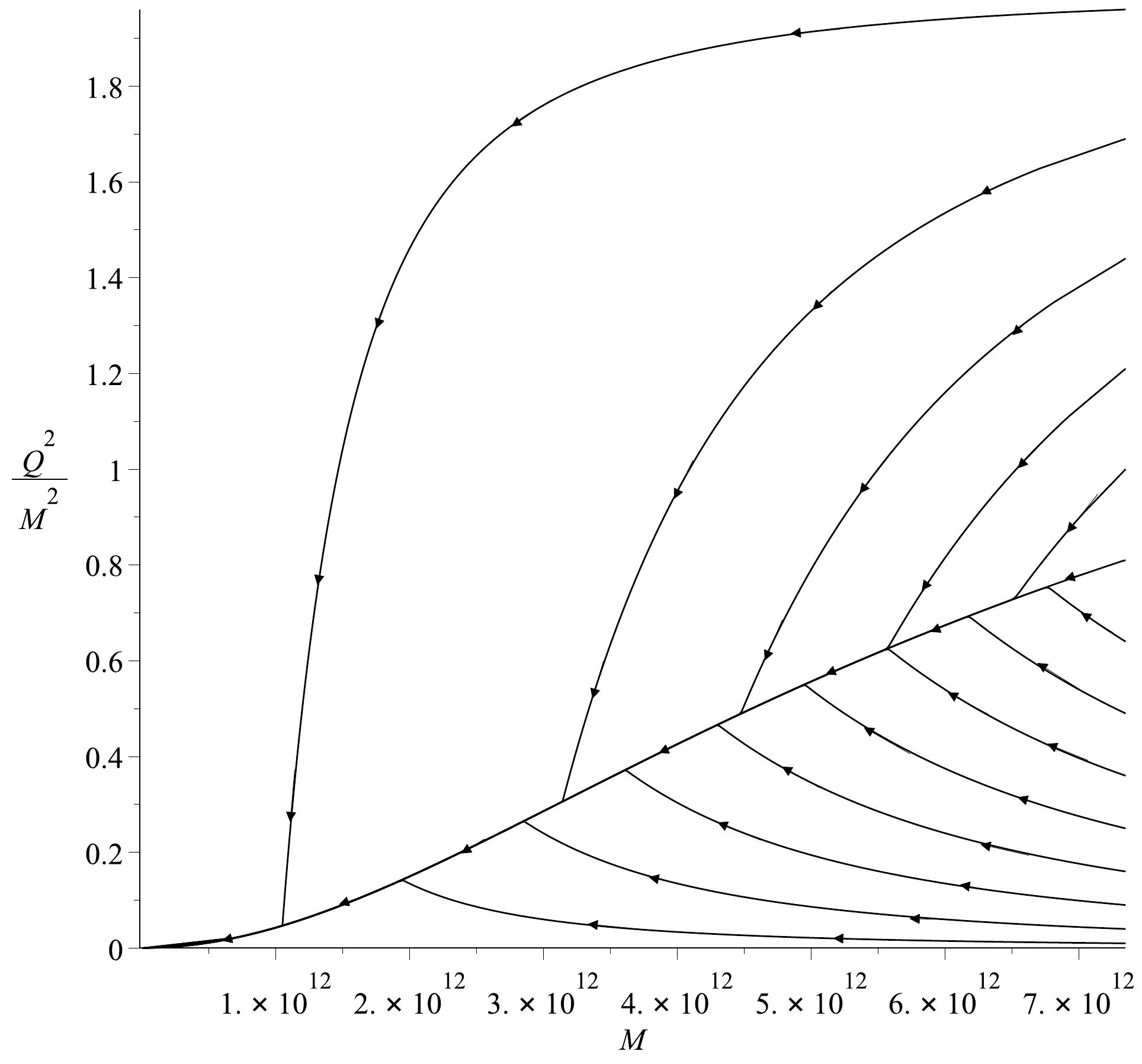}
\caption{GHS black holes avoid becoming extremal under Hiscock-Weems model, after the charged loss rate is modified by Eq.(\ref{dQdtnew}), with $\mathcal{C}=2$; c.f. Fig.(\ref{ghs-wrong-evolution}).  Arrows denote the direction of evolution.
\label{ghs-evolution}}
\end{figure}

In Fig.(\ref{ghs-evolution}), we again show the evolution of GHS black holes, but with charge loss governed by Eq.(\ref{dQdtnew}), with $\mathcal{C}=2$. We now see that evolution always lead to the Schwarzschild limit as discussed.
 
Note that in the Reissner-Nordstr\"om case we have no need to enhance the charge loss rate, since the rate is already fast enough to avoid extremality being reached.

At this point the readers might object that our modification to the charge loss rate, Eq.(\ref{dQdtnew}), is rather arbitrary or ad hoc. However, this is not the case. Remarkably one could obtain the \emph{same} modification from another consideration, which has nothing to do with cosmic censorship whatsoever (at least not directly), as was shown by Shiraishi in \cite{1305.2564}. For completeness, we will summarize the main result here, but readers should consult \cite{1305.2564} for the details.

In an attempt to understand superradience of charged dilaton black holes, Shiraishi considered a scalar field $\psi$ with mass $m$ and charge $e$. Taking the dilaton coupling into consideration, the action of the scalar field is 
\begin{equation}
S_s = \int \d{^4}x \sqrt{-g} \left[e^{-\frac{4b\phi}{3}}|(\nabla_\mu + ieA_\mu)\psi|^2 + e^{-\frac{4c\phi}{3}}m^2|\psi|^2\right],
\end{equation}
with the constraint that $b-c=1$.
Shiraishi actually considered a more general result, but we shall only focus on the GHS case. The scalar field ansatz, which allows seperation of variables, is
\begin{equation}
\psi=\frac{u(r)}{r}S_{(l)}(\theta)e^{-i\omega t},
\end{equation}
where $S_{(l)}$ is the spherical harmonic function on a 2-sphere $S^2$. To be more specific, it is the eigenfunction of the
Laplacian on $S^2$ with unit radius, such that $\nabla^2 S_{(l)}=-l(l+1)S_{(l)}$. 

Next, consider the scattering of wave on the GHS background. Shiraishi showed that in the large black hole limit, WKB approximation allows one to write the radial wave equation in the form
\begin{equation}
\frac{\d{^2} u}{\d y^2} + Wu =0,
\end{equation}
where $W=W(\omega, m, e, M, Q, b, c)$ takes a complicated form, whose details we will not repeat here. 
Here $y=y(r)$ is a rather delicate coordinate transformation defined via the ODE
\begin{equation}
\frac{\d y}{\d r} = {\left(1-\frac{Q^2}{Mr}\right)^{b-1}}{\left(1-\frac{2M}{r}\right)^{-1}}.
\end{equation}
The point is that the rate of particle emission can be calculated from $W$. First, we define
\begin{equation}
A(\omega):=\exp\left[-2\int_{y_1}^{y_2}|W|^{\frac{1}{2}} \d y\right], 
\end{equation}
where $W(y_1)=W(y_2)=0$. 
Finally, one obtains the charge loss rate by
\begin{equation}\label{shi}
\frac{\d Q}{\d t} = -e \int_m^{e\Phi_+} S_{(l)} A(\omega) \d \omega,
\end{equation}
with $\Phi_+=Q/r_+$ being the electric potential at the black hole horizon.

Shirashi showed that this formula, Eq.(\ref{shi}), when applied to the Reissner-Nordstr\"om case, recovers the Schwinger charge loss rate Eq.(\ref{dQdt}), despite in this method the emission of charged scalar particle was considered, in place of the actual leptonic electrons or positrons. Similarly for the GHS case, Shirashi shown that Eq.(\ref{shi}) yields a correction term to the charge loss rate. It turns out that this is precisely Eq.(\ref{dQdtnew}) with $\mathcal{C}=2$ that we have previously derived by imposing cosmic censorship\footnote{Shirashi's charge loss rates $\d Q/\d t$, namely Eq.(24) and Eq.(25) in \cite{1305.2564}, are actually proportional to the RHS expressions -- the numerator should be $r_+^3$ instead of $r_+$. After that, upon restoring $\hbar$ they are the same as our expression Eq.(\ref{dQdt}) and Eq.(\ref{dQdtnew}), which are manifestly dimensionless (recall that $\hbar$ has dimension of area).}.

Lastly, let us comment a little bit more on the attractor. In the work of Hiscock and Weems \cite{HW}, they claimed that the existence of attractor in the evolution of Reissner-Nordsrtr\"om black hole is due to the specific heat of these black holes changing signs depend on the values of $M$ and $Q$. While the specific heat of the RN black holes might correlate with the attractor for some subtle reason, this is not the main underlying reason. GHS black holes always have negative specific heat, just like Schwarzschild black holes. Despite this, there exists attractor behavior. In fact, the attractor essentially exists because of cosmic censorship, although admittedly it is still somewhat obscure how various physics, such as the nontrivial coupling between Maxwell field and dilaton field, conspire to uphold the censorship.

\section{Discussions: Cosmic Censorship Always Holds Under Hawking Evolution}\label{4}

While singularities are abundant in general relativity and other theories of gravitation, the hope is that singularities are generically not naked. By now it is clear from various works in the literature that naked singularities \emph{do} form under various circumstances in Einstein's gravity and beyond. The current challenge is to classify and eventually understand under which conditions do we have cosmic censorship; can one be more precise about what ``generic'' means? As emphasized in \cite{1812.05017}, in order to answer these questions, it is important to investigate various scenarios of gravitational dynamics. Exploring these issues may pave the path towards a better understanding of quantum gravity.

Since black holes evaporate via Hawking radiation, their parameter values can change throughout this process. It is conceivable that for charged black holes, $Q/M$ can evolve towards its extremal value. If black holes can become extremal via Hawking evaporation, then it is conceivable that some perturbations could render the singularities naked. Although not discussed in the context of cosmic censorship, Hiscock and Weems \cite{HW} showed that this cannot happen for asymptotically flat Reissner-Nordstr\"om black holes -- though $Q/M$ can come arbitrarily close to its extremal value for some initial conditions, it always turns away towards an attractor that in turn, flows towards the Schwarzschild limit. This is the cosmic censor in action, and can occur without any change of sign in the specific heat, contradict to the proposal by Hiscock and Weems.

The GHS black hole, which is a solution of low energy effective theory obtained from string theory, is an interesting arena to investigate cosmic censorship beyond general relativity. 
(GHS black hole with a cosmological constant has also been investigated \cite{9307177}.) After all, if string theory is correct, then strictly speaking, Reissner-Nordstr\"om black holes are not the correct charged black hole solution. Whether or not cosmic censorship still holds for the stringy solution then becomes a well-motivated question.

Various works have investigated whether naked singularities can form in this theory under gravitational collapse \cite{0209185}, or by over-charging a black hole by one way or another \cite{1803.07916, 1512.08550, 1703.07414}. Our work considers employing the scheme of Hiscock and Weems to GHS black holes to investigate whether cosmic censorship can potentially be violated under Hawking evaporation.

We showed that in order for GHS black holes to behave in the manner similar to RN black holes, one must modify the charge loss rate accordingly, so that charge particle production is slightly enhanced towards extremality. A simple argument yields Eq.(\ref{dQdtnew}) with $\mathcal{C}=2$. Remarkably, this is \emph{precisely} the form obtained by Shiraishi in \cite{1305.2564} (Eq.(25) therein) via WKB approximation method on GHS black hole background. 

The latter was done without any consideration for whether extremality can be reached\footnote{As pointed out in \cite{1305.2564}, the WKB approximation cannot be trusted very close to the extremal value. The Hiscock-Weems method also has similar limitation. However, the point is that, \emph{precisely} because of this, it is satisfying to see that evolution under Hawking evaporation will ensure that evaporating black hole naturally stays away from the extremal limit.} -- and therefore whether cosmic censorship can be violated --  under Hawking evaporation. The fact that a simple argument invoking cosmic censorship obtains this result independently lends credence to the validity of cosmic censorship in general, and in low energy effective theory of string theory specifically. 

In addition, our work also serves as an example for the \emph{utility} of the cosmic censorship. That is, by assuming that the censorship holds, we could deduce what kind of modification is required for the charge loss rate. The implication goes further than black hole physics. As explained in Sec.(\ref{2}), Hiscock and Weems derived their charge loss rate from Schwinger's formula. Although used to model Hawking radiation in some specific regime, Schwinger mechanism \emph{by itself} is of course about quantum electrodynamics, which is applicable even without the presence of gravity. Therefore this would also suggest that Schwinger mechanism gets modified in the presence of dilaton. This is of course not too surprising since the dilaton is exponentially coupled to the Maxwell field (we have assumed in this work that the Maxwell field in Einstein-Maxwell-dilaton theory, which originated from dimensional reduction of higher dimensional string theory, is the same one that coupled to the Standard Model, but this in principle need not be the case).

In a previous study by the authors, a ``complementary third law'' was found \cite{comp}, which states that if the end state of black hole evaporation has finite nonzero temperature but zero size, then the evaporation time (assuming thermal evaporation always holds) is necessarily infinite. It was shown therein that GHS black hole with electrical charge held fix, behaves in this way. It was then conjectured that allowing the charge to vary should not change this picture by much. Our current work shows that this is not true: even if we naively use Eq.(\ref{dQdt}) without the dilaton correction for the charge loss, some black holes can reach extremality in \emph{finite time} (within numerical accuracy of course), although the time scale involved is a very long one by any standard. With the correction term the behavior changes drastically and the end state is now towards complete evaporation \emph{\`a la} Schwarzschild black hole, assuming there is no new physics that might stop the evaporation near the Planck scale \cite{1412.8366}. Therefore allowing electrical charge to evaporate yields very different picture than have it fixed.

Lastly we remark that the Hawking evaporation of charged dilaton black holes could stop just prior to the formation of naked singularity due to the so-called ``mirror effect'', a timelike surface that negative energy flux does not propagate across, as shown by Maeda and Okamura in \cite{0502118}.  However, in that work, the evaporation process considered is that for five-dimensional NS1 black string with Kaluza-Klein charge (W-charge), thus it is not clear if the argument also holds for the GHS solution. If it does, this would be an extra insurance against cosmic censorship violation. 

There are two obvious questions we can ask for future investigations: Firstly, if one insists on cosmic censorship being valid, in theories in which charge loss is nontrivial (in the presence of couplings to other fields, for example), can we similarly deduce the modification one must make to the charged particle production rate? 

In order to investigate these issues, one might consider, for example, the more general EMD theory, whose action is given by
\begin{equation}
{S}[\text{EDM}(a)]= \frac{1}{16\pi}\int \text{d}^4 x \sqrt{-g}\left[R - 2(\nabla \varphi)^2 - e^{-2a\varphi}F^2\right],
\end{equation}
where $a \in \Bbb{R}^+$ is the parameter that governs the coupling strength between the dilaton and the Maxwell field.
The GHS solution corresponds to the choice $a=1$. Some aspects of the physics involved in the evaporation of such black holes have been investigated \cite{9508029v1, 1308.1877}, but it would be interesting to see if the Hiscock-Weems model would give some interesting deeper understanding, especially in the context of cosmic censorship.

The second question is more general: does cosmic censorship always hold under Hawking evaporation? \newline

\begin{acknowledgments}
YCO thanks the National Natural Science Foundation of China (No.11705162) and the Natural Science Foundation of Jiangsu Province (No.BK20170479) for funding support. He also thanks Brett McInnes for useful discussions. In addition, YCO thanks Mariam Bouhmadi L\'opez for hospitality during his visit to the University of the Basque Country UPV/EHU, Bilbao, Spain, during which this work was finalized. 
\end{acknowledgments}

\end{document}